\documentclass[aps,prd,reprint,onecolumn,10pt,floatfix,nofootinbib,
superscriptaddress,nolongbibliography]{revtex4-2}
\usepackage[T1]{fontenc}
\usepackage{lmodern,microtype}
\usepackage{amsmath,amssymb,bm,graphicx,booktabs}
\usepackage{xcolor}
\definecolor{romared}{RGB}{142,0,28}
\usepackage[unicode,colorlinks=true,linkcolor=romared,citecolor=romared,
urlcolor=romared,filecolor=romared,linktocpage,breaklinks]{hyperref}
\hypersetup{pdftitle={Infrared scalar induced gravitational waves as a probe of the cosmological equation of state}}
\newcommand{\dd}{\mathrm d}
\newcommand{\cH}{\mathcal H}
\newcommand{\Pz}{\mathcal P_\zeta}
\newcommand{\sourcecell}[2]{\parbox[t]{#1}{\raggedright #2\strut}}

\begin{document}
\title{Infrared Universality of Scalar Induced Gravitational Waves Beyond Second Order}
\author{Chen Yuan}
\affiliation{Department of Physics, College of Sciences, Shanghai University, 99 Shangda Road, 200444 Shanghai, China}
\date{September 9, 2026}
\begin{abstract}
Primordial black hole dark matter is formed from the collapse of enhanced primordial density fluctuations. The formation of primordial black holes is accompanied by scalar induced gravitational waves whose infrared scaling only depends on the equation of state, $w$, of the background at leading order. Establishing whether this characteristic scaling remains robust under corrections of arbitrary perturbative order, including in the presence of non-Gaussianities, requires a physical understanding of the infrared scaling. In this work, we provide such a physical interpretation. We find that the stress of sound waves with nearly equal frequencies produces a slow beat. Although the incoming waves oscillate rapidly, this slow beat continues to source the long wavelength tensor perturbation before its horizon entry, thus generating an infrared scaling that only depends on the expanding background. We show that higher order interactions generate freely propagating waves that finally lead to the same slow beat deep inside the horizon. Since non-Gaussianities would not change the speed of the freely propagating waves, they do not modify the infrared scaling. As the source with a slow beat has enough time to accumulate before the horizon entry of the infrared tensor mode, it records the background expansion and thus the infrared scaling could provide a robust probe of the cosmological equation of state.
\end{abstract}
\maketitle

\section{Introduction}
\label{sec:intro}
Gravitational wave (GW) astronomy \cite{Sathyaprakash:2009xs,Bailes:2021tot} has opened a new way to study both compact objects and the history of the Universe{~\cite{Bian:2025ifp}}. Beginning with GW150914 \cite{LIGOScientific:2016aoc}, observations of hundreds of compact binary coalescences have turned black hole (BH) populations into an observational subject~\cite{LIGOScientific:2018mvr,LIGOScientific:2020ibl,KAGRA:2021vkt,LIGOScientific:2025slb,LIGOScientific:2026wfs}. These events have also renewed interest in primordial black holes (PBHs) in explaining the origin of these compact objects. Primordial interpretations were proposed for the first binary BH events {\cite{Bird:2016dcv,Sasaki:2016jop,Clesse:2016vqa}}, and continue to be investigated for more recent systems \cite{LIGOScientific:2024elc,Huang:2024wse,Nakamura:1997sm,Ali-Haimoud:2017rtz,Raidal:2018bbj,Vaskonen:2019jpv,Chen:2021nxo,LIGOScientific:2020iuh,LIGOScientific:2020zkf,Prunier:2023uoo,Yuan:2024yyo,Yuan:2025avq}.

PBHs are BHs formed in the early Universe and are promising dark matter candidates~\cite{Zeldovich:1967lct,Hawking:1971ei,Carr:1974nx,Carr:1975qj}. We consider their standard formation channel where a sufficiently large primordial density fluctuation collapses after its horizon entry. The contribution of PBHs to dark matter is constrained by several independent observations. See, e.g., ~\cite{Carr:2020gox,Green:2020jor} for reviews of PBH constraints. To explain most of the dark matter in the universe, the power spectrum of the curvature perturbations needs to be enhanced by several orders of magnitude over the cosmic microwave background (CMB) scales \cite{Sasaki:2018dmp,Green:2020jor}. Such an enhancement can arise from features in the inflationary potential~\cite{Ivanov:1994pa,Garcia-Bellido:2017mdw,Ballesteros:2017fsr}, transitions involving several fields~\cite{Garcia-Bellido:1996mdl,Inomata:2017okj,Pi:2017gih}, resonant amplification~\cite{Cai:2018tuh,Cai:2019bmk}, Higgs related dynamics~\cite{Ezquiaga:2017fvi,Espinosa:2017sgp} or non-minimal spectator models~\cite{Meng:2022low}.

During the formation of PBHs, linear order scalar perturbations couple together and source the higher order metric perturbations{~\cite{Mollerach:2003nq,Ananda:2006af,Baumann:2007zm}}. The propagating part is known as the scalar induced gravitational waves (SIGWs)~\cite{Domenech:2020xin,Yuan:2025seu,Domenech:2017ems,Inomata:2019yww,Yuan:2019fwv,Yuan:2025seu}. For reviews of SIGWs, see e.g., \cite{Yuan:2021qgz,Domenech:2021ztg}. {The phenomenology of SIGWs spans several frequency ranges. Pulsar timing arrays (PTAs) probe the nanohertz band, where recent data provide evidence for a correlated GW background~\cite{NANOGrav:2023gor,EPTA:2023fyk,Reardon:2023gzh,Xu:2023wog}. Searches and interpretations using SIGW templates connect the PTA data with the PBH scenarios~\cite{Chen:2019xse,NANOGrav:2023hvm,Liu:2023pau,Chen:2024twp,Liu:2023ymk}. At millihertz frequencies, the Laser Interferometer Space Antenna (LISA)~\cite{LISA:2017pwj} and Taiji~\cite{Hu:2017mde} give forecasts of asteroid mass PBHs through SIGWs~\cite{Bartolo:2018evs,Jiang:2026krg} and the reconstruction of the primordial scalar spectrum~\cite{LISACosmologyWorkingGroup:2024hsc,LISACosmologyWorkingGroup:2025vdz}. Moreover, searches of SIGWs in the LIGO/Virgo/KAGRA data provide constraints on ultralight PBHs~\cite{Jiang:2024aju}. 

The energy spectrum of SIGWs depends on the primordial curvature power spectrum. However, small scale primordial power spectrum is much less constrained than that on CMB scales, leading to difficulties in distinguishing the SIGW signal from other GW signals. Fortunately, the SIGWs from PBHs are supposed to have characteristic features in their infrared spectrum. At wavelengths much longer than those of the enhanced scalar perturbations, the scaling of the SIGW spectrum in radiation domination (RD) is approximately $k^3\ln^2(k_*/k)$ \cite{Yuan:2019wwo,Pi:2020otn,Li:2024lxx,Yuan:2023ofl,Chen:2026hnv}. A sufficiently narrow feature can also display an intermediate $k^2\ln^2(k_*/k)$ scaling~\cite{Yuan:2019wwo,Pi:2020otn}. For a constant background equation of state $w$, the corresponding infrared scaling only depends on $w$ \cite{Domenech:2019quo,Domenech:2020kqm}. Moreover, even in the presence of primordial non-Gaussianities, the infrared scaling of SIGWs is still model independent \cite{Cai:2018dig,Yuan:2023ofl}. The infrared scaling is then thought to be a unique feature of SIGWs and is used to constrain the early equation of state \cite{Liu:2023pau,Domenech:2024rks,Hajkarim:2019nbx,Abe:2020sqb,Escriva:2024ivo,Liu:2023hpw}. 

However, most of the studies concerning the infrared scaling focus on the leading order of SIGWs. Although the calculations of SIGW have been extended to third order~\cite{Yuan:2019udt,Chen:2019xse,Zhou:2021vcw,Chang:2022nzu,Wang:2023sij,Chang:2023vjk,Zhao:2026jax}, these papers do not provide a full third order result. A full third order correction to the SIGW spectrum contains both the third order tensor autocorrelation $\langle h_3h_3\rangle$ and the second and fourth order cross correlation $\langle h_2h_4\rangle$. \cite{Yuan:2019udt} includes both types of correlation, but only consider first order scalar perturbations in the calculation. Subsequent studies include the induced second order scalar, vector, and tensor perturbations but only evaluate the autocorrelation of $h_3$~\cite{Zhou:2021vcw,Chang:2022nzu}. The corresponding $\langle h_2h_4\rangle$ contribution with the full impact from induced perturbations has not been included in these calculations. Whether the characteristic infrared scaling remains robust at arbitrary perturbative order is therefore still an open question. In this paper, we focus on the physical origin of this scaling and explore the infrared scaling under higher order corrections.

This paper is organized as follows. Sec.~\ref{sec:second} introduces the metric perturbations up to second order and gives a physical interpretation of the second order infrared scaling. Sec.~\ref{sec:third} generalizes this interpretation up to third order diagrams. Sec.~\ref{sec:all} investigates the impact of arbitrary higher order contributions. Sec.~\ref{sec:nongaussian} discusses non-Gaussian perturbations and their implications for the infrared scaling. {Sec.~\ref{sec:conclusions} summarizes our conclusions}. Detailed derivations are given in the appendices.

\section{The slow beat of the Second order SIGWs}
\label{sec:second}
\subsection{Metric, fluid, and first order field equations}
We use the framework of cosmological perturbation theory beyond linear order in Newtonian gauge~\cite{Matarrese:1997ay,Noh:2003yg,Nakamura:2004rm,Malik:2008im} and consider an adiabatic perfect fluid with constant equation of state $0<w\leq1/3$ and sound speed $c_s^2=w$. The scalar, vector, and tensor parts of the metric are given by
\begin{equation}
 \begin{aligned}
 \dd s^2&=a^2(\eta)\left\{-(1+2\Phi)\dd\eta^2
 +2V_i\dd\eta\dd x^i+
 \left[(1-2\Psi)\delta_{ij}+h_{ij}\right]\dd x^i\dd x^j\right\},\\
 V_i&=\sum_{n\geq2}V_{ni},\quad \partial_iV_{ni}=0,
 \qquad h_{ij}=\sum_{n\geq2}h_{n,ij},\quad
 \partial_i h_{n,ij}=h_{n,ii}=0.
 \end{aligned}
 \label{eq:metric}
\end{equation}
Here, $\Phi$ and $\Psi$ are the Bardeen potentials and their higher order components are defined as $\Phi=\Phi_1+\Phi_2+\cdots$ and $\Psi=\Psi_1+\Psi_2+\cdots$. We also consider the higher order tensor modes $h_{n,ij}$, which satisfy the transverse and traceless (TT) condition and higher order vector modes $V_i$, which satisfy the transverse condition. Throughout this paper, we only consider the initial value of the first order scalar modes, as they are much stronger than other modes during the formation of PBHs.

We take the first order initial curvature perturbation to be Gaussian. This assumption is relaxed in Sec.~\ref{sec:nongaussian}. The energy momentum tensor and fluid variables are
\begin{align}
 T_{\mu\nu}&=(\rho+P)U_\mu U_\nu+Pg_{\mu\nu},
 \qquad\rho=\bar\rho+\delta\rho_1+\cdots,\qquad
 P=\bar P+\delta P_1+\cdots,\qquad
 \bar P=w\bar\rho,\quad\delta P_1=w\delta\rho_1.
 \label{eq:fluid}
\end{align}
where $\rho,P,U^\mu$ are the energy density, pressure and four velocity of the fluid. Bars denote background quantities, and the $\delta$ symbol denotes the perturbations of the fluid.
We expand the four velocity as $aU^i=\mathcal V_1^i+\mathcal V_2^i+\cdots$ so that $\mathcal V_n^i$ is the $n$-th order velocity perturbation. $\eta$ is the conformal time and a prime denotes its derivative. The flat background satisfies
\begin{equation}
 b=\frac{1-3w}{1+3w},\qquad 0\leq b<1,\qquad
 a\propto\eta^{1+b},\qquad \cH\equiv a'/a=\frac{1+b}{\eta},
 \qquad 3\cH^2=8\pi G a^2\bar\rho.
 \label{eq:bg}
\end{equation}

The first order Einstein equations can be written as:
\begin{align}
 \nabla^2\Phi_1-3\cH(\Phi_1'+\cH\Phi_1)
 &=4\pi Ga^2\delta\rho_1, &&(00),\label{eq:einstein00}\\
 \partial_i(\Phi_1'+\cH\Phi_1)
 &=-4\pi Ga^2(\bar\rho+\bar P)\mathcal V_{1i},
 &&(0i),\label{eq:einstein0i}\\
 \Phi_1''+3\cH\Phi_1'+(2\cH'+\cH^2)\Phi_1
 &=4\pi Ga^2\delta P_1, &&(ij\ \hbox{trace}),
 \label{eq:einsteinij}
\end{align}
where we have used the relation $\Psi_1=\Phi_1$ due to the ij component of the first order Einstein equation.
With the above equations, the equation of motion for $\Phi_1$ can be obtained as:
\begin{equation}
  \Phi_1''+\frac{4+2b}{\eta}\Phi_1'+c_s^2p^2\Phi_1=0.
\label{eq:linear}
\end{equation}
This is a damped sound wave equation. We define the initial value as  $\Phi_1(\bm p,0)=3(1+w)\zeta(\bm p)/(5+3w)$ where $\zeta$ is the comoving curvature perturbation.
Writing $\Phi_1(\bm p,\eta)=T_w(p,\eta)\zeta(\bm p)$ separates the initial amplitude from its subsequent evolution. The regular transfer function $T_w(p,\eta)$ is given by
\begin{equation}
 T_w(p,\eta)=\frac{3(1+w)}{5+3w}\,
 2^\alpha\Gamma(\alpha+1)
 \frac{J_\alpha(c_sp\eta)}{(c_sp\eta)^\alpha},
  \qquad \alpha=b+\frac32.
\label{eq:regulartransfer}
\end{equation}
Here $J_\alpha$ is the ordinary Bessel function of the first kind. Well inside the sound horizon, $c_sp\eta\gg1$, its asymptotic behavior is
\begin{align}
 T_w(p,\eta)&=\mathcal N_w(c_sp\eta)^{-2-b}
 \cos\left[c_sp\eta-\frac{\pi(b+2)}2\right]
 +\mathcal{O}((c_sp\eta)^{-3-b}),\qquad
 \mathcal N_w=\frac{3(1+w)}{5+3w}
 \frac{2^{b+2}\Gamma(b+5/2)}{\sqrt\pi}.
 \label{eq:potentialenvelope}
\end{align}
The Bardeen potential oscillates rapidly at a frequency of $c_s p$ inside a decaying envelope. The velocity behaves differently. In Fourier space, the velocity perturbation behaves as
\begin{equation}
 \mathcal V_{1i}(\bm p,\eta)=
 \frac{2i\widehat p_i\mathcal N_w}{3(1+w)c_s(1+b)^2}
 (c_sp\eta)^{-b}
 \sin\left[c_sp\eta-\frac{\pi(b+2)}2\right]\zeta(\bm p)
 +\mathcal{O}((c_sp\eta)^{-1-b})\zeta(\bm p),
 \label{eq:velocityenvelope}
\end{equation}
in the subhorizon limit. The envelopes are therefore $\Phi_1\propto\eta^{-2-b}$ and $\mathcal V_1\propto\eta^{-b}$. In RD the Bardeen potential decreases as $\eta^{-2}$, whereas the leading velocity oscillation has constant amplitude. We will show in the next subsection that the velocity perturbation is the one that contributes to the infrared scaling of SIGWs.

\subsection{The slow beat generated by velocity perturbations and the infrared tensor modes}

\begin{figure}[htb]
\centering
\includegraphics[trim=2pt 2pt 2pt 2pt,clip]{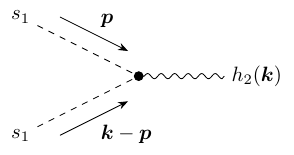}
\caption{Diagram of the second order tensor mode, $h_2$, induced by two first order scalar perturbations, $s_1$.}
\label{fig:second}
\end{figure}

Expanding the spatial Einstein equation and retaining its TT part gives
\begin{equation}
 h_{2,ij}''+2\cH h_{2,ij}'-\nabla^2h_{2,ij}
 =S_{2,ij},\qquad
 S_{2,ij}=\Lambda_{ij}{}^{ab}
 \left[4\,\partial_a\Phi_1\partial_b\Phi_1
 +16\pi Ga^2(\bar\rho+\bar P)\mathcal V_{1a}\mathcal V_{1b}\right].
 \label{eq:physicalsource}
\end{equation}
Here $\Lambda$ is the TT projection operator. For polarization tensors normalized by $e_\lambda^{ij}e^*_{ij,\lambda'}=\delta_{\lambda\lambda'}$, define $h_{2,\lambda}=e_\lambda^{ij}h_{2,ij}$ and similarly for $S_{2,\lambda}$. The solution of $h_2$ can be obtained using the Green's function:
\begin{align}
 h_{2,\lambda}(\bm k,\eta)
 &=\int_{\eta_i}^{\eta}\dd\tau\,
 G_k(\eta,\tau)S_{2,\lambda}(\bm k,\tau),\label{eq:retardedh}\\
 (\partial_\eta^2+2\cH\partial_\eta+k^2)G_k(\eta,\tau)
 &=\delta(\eta-\tau),\qquad G_k(\eta,\tau)=0\quad(\eta<\tau).
 \label{eq:tensorgreen}
\end{align}
Here, $\eta_i$ is the initial time and $G_k$ is the Green's function for the tensor mode. In Fourier space the velocity stress is a convolution,
\begin{equation}
  S_{2,\lambda}^{\mathcal V}(\bm k,\eta)
 =16\pi Ga^2(\bar\rho+\bar P)
 \int\frac{\dd^3p}{(2\pi)^3}e_\lambda^{ij}(\bm k)
\mathcal V_{1i}(\bm p,\eta)
  \mathcal V_{1j}(\bm k-\bm p,\eta).
 \label{eq:velocitysource}
\end{equation}
For the infrared regime of SIGWs, we need to focus on $k\ll k_*$ where $k_*\in[k_-,k_+]$ is a pivot scale of the enhanced power spectrum. This indicates the frequencies of the two scalar modes in Fig.~\ref{fig:second} are nearly equal.  Both incoming waves in Fig.~\ref{fig:second} can remain in the same enhanced band as $k$ decreases in the infrared regime, producing a long wavelength which is far away from the enhanced band. We will demonstrate below that this long wavelength is a slow beat that contributes to the infrared regime of SIGWs.

For convenience, we define
\begin{equation}
  \begin{gathered}
 u\equiv\frac{|\bm p-\bm k|}{k_*},\qquad
 v\equiv\frac{p}{k_*},\qquad \tilde{k}\equiv\frac{k}{k_*},\qquad
 \mu=\widehat{\bm k}\cdot\widehat{\bm p}, \qquad
 u=\sqrt{v^2+\tilde{k}^2-2v\tilde{k}\mu}
 =v-\tilde{k}\mu+O(\tilde{k}^2/v)
 \end{gathered}
 \label{eq:pair}
\end{equation}
Since the velocity perturbation is described by a sine function with phase $\varphi=-\pi(b+2)/2$, the second order source in Eq.~(\ref{eq:velocitysource}) should contain:
\begin{equation}
  \sin(c_svk_*\eta+\varphi)\sin(c_suk_*\eta+\varphi)
 =\frac12\cos[c_sk_*(v-u)\eta]
 -\frac12\cos[c_sk_*(v+u)\eta+2\varphi].
 \label{eq:secondtrig}
\end{equation}
This is a slow modulation of two rapid sound waves, namely a beat of the acoustic wave. The slow beat frequency is $\sim\mathcal{O}(k)$,
\begin{equation}
  \Delta\omega=c_sk_*(v-u)=c_sk\mu+O(c_sk^2/p)\sim \mathcal{O}(k).
\label{eq:slowfreq}
\end{equation}
On the other hand, Eq.~(\ref{eq:velocityenvelope}) implies the velocity contributes as $\eta^{-b}$. Since $16\pi Ga^2(\bar\rho+\bar P)=6(1+w)\cH^2\propto\eta^{-2}$, the envelope of Eq.~(\ref{eq:velocitysource}) is thus $\eta^{-2-2b}$.  Combining the envelope and the beat frequency gives
\begin{equation}
S_{2,\lambda}^{\mathcal V}(\bm k,\eta)
   \sim\eta^{-2-2b}
 \int\frac{\dd^3p}{(2\pi)^3}
 B_{\lambda}\cos[c_sk_*(v-u)\eta] + \text{high frequency term}.
\label{eq:beatsource}
\end{equation}
Here $B_\lambda$ is independent of $\eta$ which denotes the coefficient from the two initial $\zeta$ amplitudes, polarization factors and the normalization factors. Its explicit expression is given in Appendix~\ref{app:second}.
We choose $\eta_s$ with $c_sk_-\eta_s\gg1$ and $k\eta_s\ll1$ as a reference time when the enhanced scalar modes already oscillate as sound waves while the tensor mode remains outside the horizon, marking the start of the accumulation. Before the horizon entry of the tensor mode, $|c_s k_*(v-u)\eta|\ll1$ so that the cosine term approaches a constant. Hence, the slow beat contribution is
\begin{equation}
    S_{2,\lambda}^{\mathcal{V}}(\eta)
    \supset
    S_s
    \left(\frac{\eta}{\eta_s}\right)^{-2-2b}
    \label{eq:ir-beat-source}
\end{equation}
Here $S_s$ contains the time independent factors.
We can now evaluate the infrared behavior from Eq.~(\ref{eq:retardedh}) and Eq.~(\ref{eq:beatsource}). Before tensor horizon entry, $k\eta\ll1$, the result can be obtained as
\begin{equation}
\begin{aligned}
    h_{2}(\eta)
    &\simeq
    \int_{\eta_s}^{\eta}
    d\tau\,G_0(\eta,\tau)S(\tau)
    =
    \frac{S_s\eta_s^{2+2b}}{1+2b}
    \int_{\eta_s}^{\eta}
    \frac{d\tau}{\tau^{1+2b}}
    \left[
        1-\left(\frac{\tau}{\eta}\right)^{1+2b}
    \right]
    \label{eq:ir-beat-response}
\end{aligned}
\end{equation}
The remaining time integral gives
\begin{equation}
    h_{2}(\eta)
    \simeq
    S_s\eta_s^2
    \begin{cases}
        \displaystyle
        \frac{
            1-(1+2b)(\eta/\eta_s)^{-2b}
            +2b(\eta/\eta_s)^{-1-2b}
        }{
            2b(1+2b)
        },
        & 0<b<1,
        \\[3mm]
        \displaystyle
        \ln(\eta/\eta_s)-1+(\eta/\eta_s)^{-1},
        & b=0.
    \end{cases}
    \label{eq:ir-explicit-time-response}
\end{equation}
For any fixed $b>0$, this contribution approaches the constant $S_s\eta_s^2/[2b(1+2b)]$ as $\eta/\eta_s$ increases. In RD, $b=0$, it instead grows logarithmically. Tensor horizon entry at $\eta_k\sim k^{-1}$ sets the end of this accumulation. Consequently, RD produces an amplitude proportional to $\ln[1/(k\eta_s)]\simeq\ln(k_*/k)$, while the amplitude approaches a constant for $b>0$.

The GW energy density per logarithmic wavenumber
is defined by
\begin{equation}
    \Omega_{\rm GW}(k,\eta)
    \equiv
    \frac{1}{\rho_c(\eta)}
    \frac{d\rho_{\rm GW}(\eta)}{d\ln k}=
    \frac{1}{12}
    \left(\frac{k}{\mathcal{H}}\right)^2\frac{k^3}{2\pi^2}
    \overline{\left\langle
    h_{ij}(\bm k,\eta)h^{ij}(-\bm k,\eta)
    \right\rangle'},
\label{eq:ir-omega-definition}
\end{equation}
where $\rho_c$ is the critical energy density of the universe. Here, a prime on the two point function means neglect the factor $(2\pi)^3\delta^3(\bm k+\bm k')$. The overline denotes an average over the tensor oscillations, and the last equality applies to freely propagating tensors well inside the horizon. Details of the two point function and power spectrum are given in  ppendix~\ref{app:second}.  Due to the slow beat of the sound waves, the spectrum of SIGWs is thus
\begin{equation}
    \Omega_{22}(k,\eta_f)
    \propto
    \begin{cases}
        \displaystyle
        k^{3-2b},
        & 0<w<1/3,
        \\[2mm]
        \displaystyle
        k^3
        \ln^2(k_*/k),
        & w=1/3.
    \end{cases}
    \label{eq:ir-second-order-spectrum}
\end{equation}

To illustrate Eq.~(\ref{eq:ir-second-order-spectrum}), Fig.~\ref{fig:rd-numerical} shows the second order energy spectrum of SIGWs during RD for two kinds of enhanced curvature spectra: log-normal (LN) and broken power-law (BPL),
\begin{equation}
 \mathcal P_\zeta^{\rm LN}(p)
 =A\exp\!\left[-\frac{\ln^2(p/k_*)}{2\sigma^2}\right],
\end{equation}
\begin{equation}
 \mathcal P_\zeta^{\rm BPL}(p)
 =\frac{2A}{(p/k_*)^{-\beta}+(p/k_*)^{\beta}}.
\end{equation}
We fix $\sigma=0.2,0.5,0.8$ and $\beta = 2,4,6$.
As shown in the figure, all these power spectrum yield the same infrared scaling $k^3\ln^2(k_*/k)$.

\begin{figure}[t]
\centering
\includegraphics[width=0.8\linewidth]{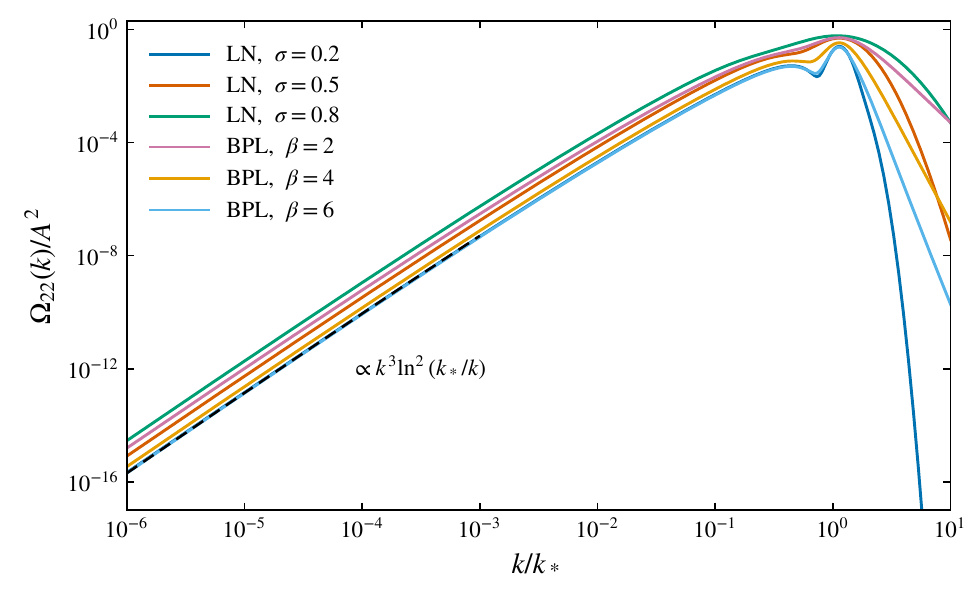}
\caption{The energy spectrum of second order SIGW during RD for the log-normal
and broken power-law curvature spectra. The black dashed curve denotes the scaling 
$k^3\ln^2(k_*/k)$.}
\label{fig:rd-numerical}
\end{figure}

Now, the physical picture of the second order SIGWs is clear. Firstly, the scalar waves are already oscillating by the reference time $\eta_s$, but the tensor mode in the infrared regime enters the horizon much later, at $\eta_k\sim k^{-1}$. Throughout $\eta_s\ll\eta\lesssim\eta_k$, the scalar waves oscillate rapidly, while their product generates a slow beat. This beat frequency is so slow that its oscillation can be ignored, causing the source term to decay according to its envelope. The tensor modes therefore receive a continuous source from the scalar waves that are deep inside the horizon. As a result, the leading infrared index is nearly independent of the detailed shape of the enhanced spectrum in $[k_-,k_+]$ and only depends on the equation of state $w$. The infrared scaling can therefore serve as a probe of the cosmological background. The physical mechanism is illustrated schematically in Fig.~\ref{fig:slow-beat-mechanism}.

\begin{figure}[t]
\centering
\includegraphics[width=\linewidth]{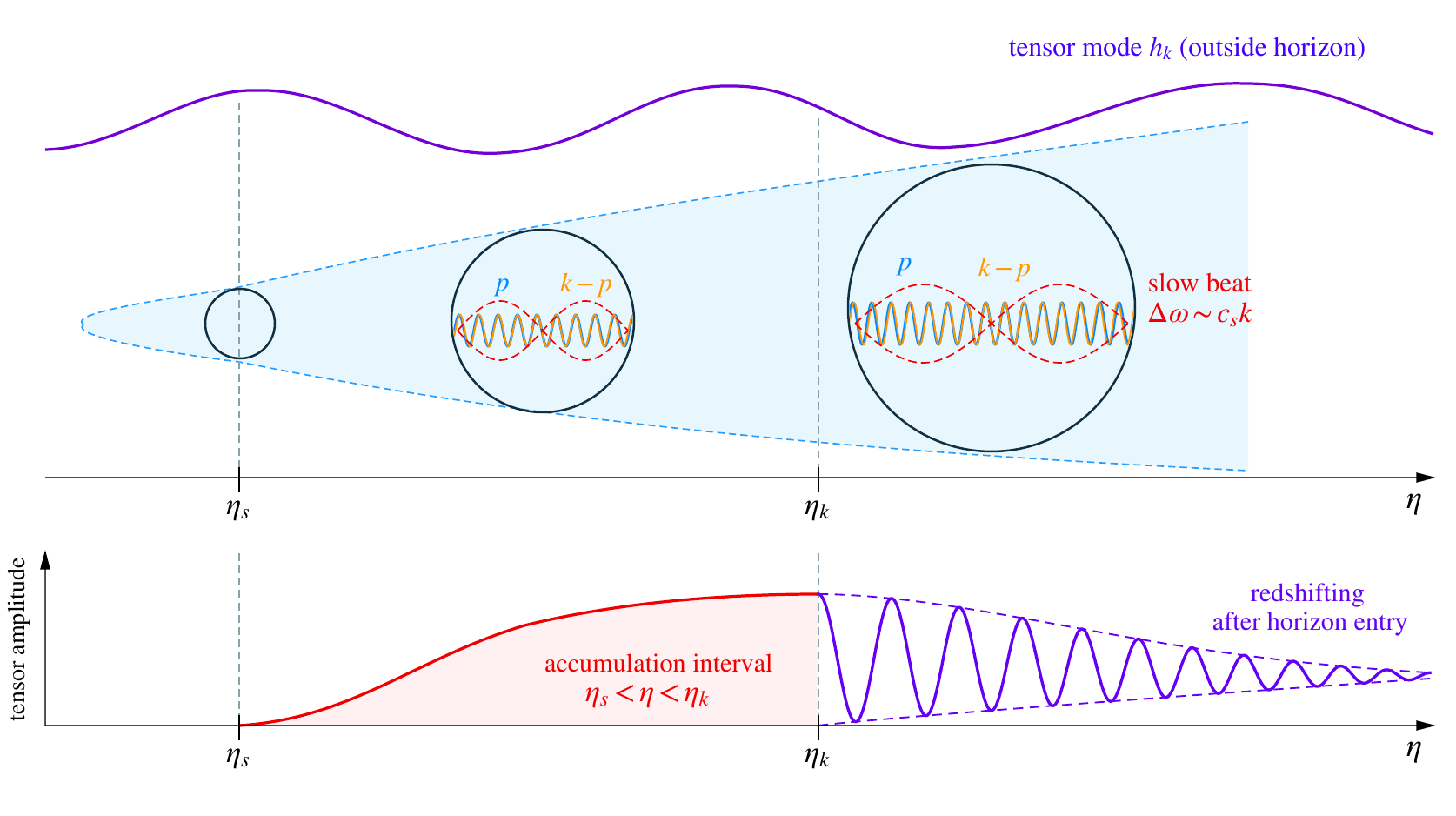}
\caption{Schematic illustration of the slow beat mechanism. Upper panel: scalar waves $\bm p$ and $\bm k-\bm p$ oscillate rapidly and produce a slow beat with a frequency $\Delta\omega\sim c_s k$. Lower panel: the tensor response accumulates during $\eta_s<\eta<\eta_k=k^{-1}$ and redshifting after horizon entry. This schematic was made with the assistance of OpenAI GPT-6 Astra under the author's direction.}
\label{fig:slow-beat-mechanism}
\end{figure}

For completeness, we also show that the other second order source contributions do not provide a stronger infrared scaling in Appendix~\ref{app:second}. Therefore, the infrared scaling arises from the slow beat of the source.

\section{The slow beat of the Third order SIGWs}
\label{sec:third}
\subsection{Two classes of source}
As we discussed above, the infrared scaling could be independent of the detailed spectrum in the enhanced regime $[k_-,k_+]$ and only depends on the cosmological background. In this section, we explore this scaling for the third order SIGWs.
There are only two ways to form a third order source. The first occurs when three linear scalar perturbations interact directly (labeled as the $1+1+1$ diagram). The second occurs when a linear scalar perturbation interacts with a second order scalar, vector, or tensor perturbation (the $1+2$ diagram). These possibilities give the four diagrams
in Fig.~\ref{fig:third}.

In the next few subsections, we will demonstrate that the scalar $1+2$ diagram, panel (b), yields a slow beat throughout the convolution. It explains the leading RD logarithm scaling. Other sources are discussed in the appendices. For the third order convolution, we define the following dimensionless momentum ratios for convenience:
\begin{equation}
p=v k_*,\quad |\bm k-\bm p|=u k_*,\quad
  q=r k_*,\quad |\bm p-\bm q|=\ell k_*.
\label{eq:thirdrouting}
\end{equation}
Thus $u,v,r,\ell$ are dimensionless momentum ratios.
In a $1+2$ diagram, the first two scalar modes generate an intermediate mode with wavelength $\bm p$, which is then convolved with the third mode, with wavelength $\bm{k-p}$, as shown in Fig.~\ref{fig:third}.

\begin{figure}[htb]
\centering
\includegraphics[trim=2pt 2pt 2pt 2pt,clip,width=\linewidth]{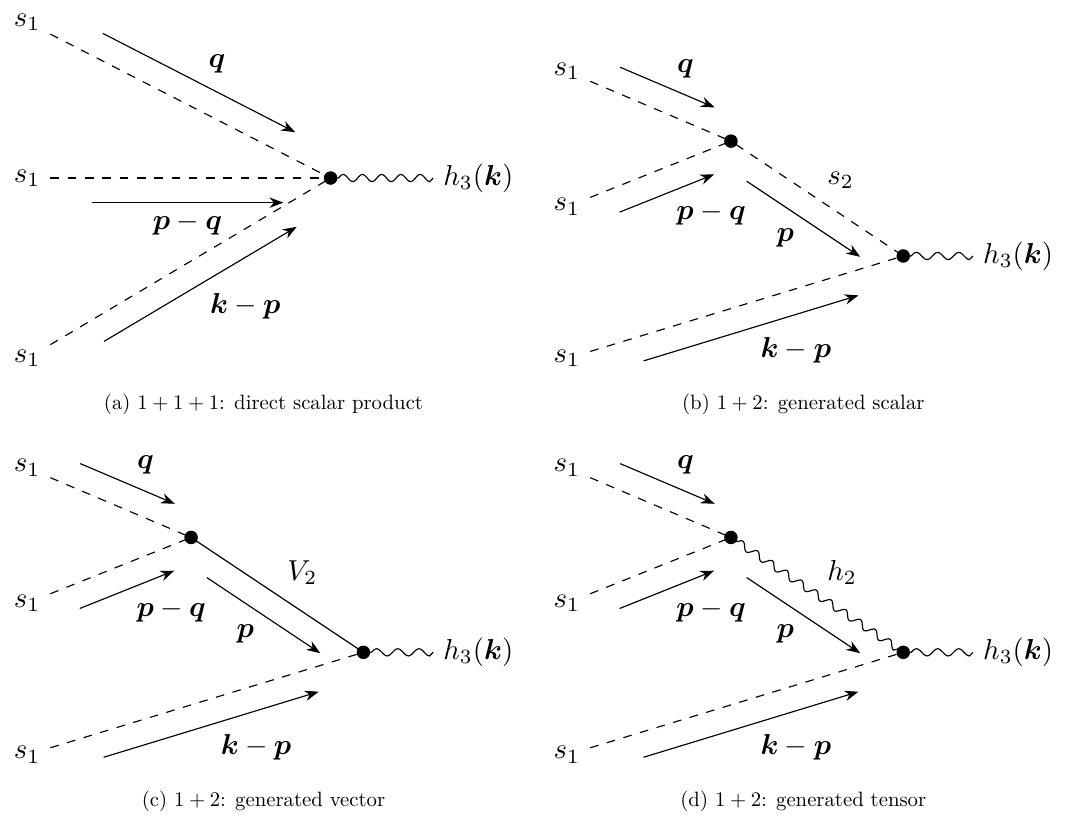}
\caption{All possible diagrams of the third order tensor mode $h_3$. Dashed, solid, and wavy lines denote scalar, vector, and tensor modes.}
\label{fig:third}
\end{figure}

\subsection{The slow beat caused by the scalar \texorpdfstring{$1+2$}{1+2} diagram}
The first interaction in Fig.~\ref{fig:third}(b) combines two first order scalar waves into a second order scalar mode. The equation of motion for the spatial Bardeen potential is
\begin{equation}
 \Psi_2''(\bm p)+\frac{4+2b}{\eta}\Psi_2'(\bm p)+c_s^2p^2\Psi_2(\bm p)
 =\int\frac{\dd^3q}{(2\pi)^3}
 Q_w(\bm p,\bm q;\eta)\zeta(\bm q)\zeta(\bm p-\bm q),
 \label{eq:scalarforced}
\end{equation}
where $Q_w$ is the quadratic source made from the first order Bardeen potential and its derivatives. In Appendix~\ref{app:scalar}, we derive its explicit expression for constant $w$. Eq.~(\ref{eq:scalarforced}) is a forced wave equation. This equation can be solved using the Green's function:
\begin{equation}
 \Psi_2(\bm p,\eta)=
 \int\frac{\dd^3q}{(2\pi)^3}\zeta(\bm q)\zeta(\bm p-\bm q)
\int_{\eta_i}^{\eta}\dd\tau\,
  G_\Psi(p;\eta,\tau)Q_w(\bm p,\bm q;\tau).
 \label{eq:scalarretarded}
\end{equation} 
The two waves entering the source $Q_w$ have wavelengths $\bm q$ and $\bm p-\bm q$ and frequencies $c_sk_*r$ and $c_sk_*\ell$. Their product drives oscillations at $c_sk_*(r+\ell)$ and $c_sk_*|r-\ell|$. On the other hand, the homogeneous equation describes free propagation at frequency $c_sp=c_sk_*v$. This can be seen from the Green's function:
\begin{equation}
 G_\Psi(p;\eta,\tau)\simeq
 \left(\frac{\tau}{\eta}\right)^{2+b}
\frac{\sin[c_sp(\eta-\tau)]}{c_sp},
 \qquad c_sp\eta,\ c_sp\tau\gg1.
 \label{eq:scalaracousticgreen}
\end{equation}
The sine function describes a sound wave with frequency $c_sp$ and the factor describes its damping by the expanding background. The full expression of the Green's function is given in Appendix~\ref{app:scalar}. 
We next determine the velocity perturbation carried by this generated scalar mode.
The scalar part of the second order $0i$ Einstein equation gives
\begin{equation}
  \mathcal V_{2i}(\bm p,\eta)
 =-\frac{2ip_i}{3(1+w)\cH^2}
       [\Psi_2'+\cH\Phi_2](\bm p,\eta)
       +\mathcal R_{2i}(\bm p,\eta).
\label{eq:velocityconstraint}
\end{equation}
Here, $\mathcal R_{2i}$ contains products of first order fields and its explicit expression is derived with $Q_w$ in Appendix~\ref{app:scalar}.
The velocity therefore contains the forced frequencies $c_sk_*(r+\ell)$ and $c_sk_*|r-\ell|$, together with the freely propagating frequency $c_sk_*v$.
For the freely propagating part, the leading derivative gives
$\Psi_2'\sim c_sp\Psi_2$, while $\cH^{-2}\propto\eta^2$.
Consequently,
\begin{equation}
 \Psi_2^{\rm prop}\propto\eta^{-2-b},\qquad
 \mathcal V_2^{\rm prop}\propto\eta^2\Psi_2^{\rm prop}
 \propto\eta^{-b}.
 \label{eq:thirdvelocityenvelope}
\end{equation}
The second order sound wave thus carries a velocity with the same time envelope as the first order wave in Eq.~(\ref{eq:velocityenvelope}). This is as expected since the homogeneous part of Eq.~(\ref{eq:scalarforced}) and the relation between $\Psi_2^{\text{prop}}$ and $\mathcal{V}_2^{\text{prop}}$ are the same as those at first order.

The vertex in Fig.~\ref{fig:third}(b) now couples these two velocities. The contribution to the third order tensor source is thus
\begin{equation}
 \begin{aligned}
 S_{3,\lambda}^{\mathcal V,12}(\bm k,\eta)
 ={}&16\pi Ga^2(\bar\rho+\bar P)
 \int\frac{\dd^3p}{(2\pi)^3}e_\lambda^{ij}(\bm k)\left[
 \mathcal V_{1i}(\bm k-\bm p,\eta)\mathcal V_{2j}(\bm p,\eta)
 +\mathcal V_{2i}(\bm p,\eta)\mathcal V_{1j}(\bm k-\bm p,\eta)
 \right].
 \end{aligned}
 \label{eq:thirdstress}
\end{equation}
This is similar to Eq.~(\ref{eq:velocitysource}) where one of the first order velocities is replaced by $\mathcal{V}_2$. This source term includes the convolution of the freely propagating part, $\mathcal{V}_2^{\text{prop}}$ (with frequency $c_sk_*v$) and the first order part {$\mathcal{V}_1$} (with frequency $c_sk_*u$). As we have already shown in Sec.~\ref{sec:second}, such a convolution yields a slow beat frequency $\Delta\omega = c_s k_* (v-u)$, namely
\begin{equation}
 \begin{aligned}
  S_{3,\lambda}^{\rm beat}(\bm k,\eta)
 \propto \left(\frac{\eta}{\eta_s}\right)^{-2-2b}
\int\frac{\dd^3p}{(2\pi)^3}
  \cos(\Delta\omega\eta).
\end{aligned}
 \label{eq:thirdsource}
\end{equation}
The cosine beat has the same $\eta^{-2-2b}$ envelope and the same $\mathcal{O}(k)$ frequency as the second order source. Before the infrared tensor enters the horizon, $|\Delta\omega\eta|\ll1$, the cosine approaches a constant. The Green's function integral for $h_3$ therefore yields the same form of Eq.~(\ref{eq:ir-beat-response}) and Eq.~(\ref{eq:ir-explicit-time-response}).

\begin{figure}[htb]
\centering
\includegraphics[trim=2pt 2pt 2pt 2pt,clip,width=.60\linewidth]{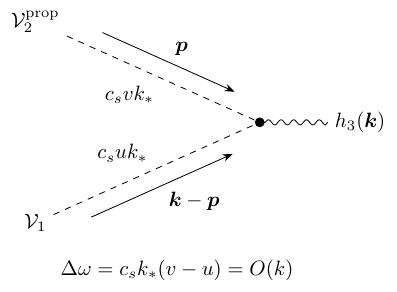}
\caption{The last interaction of the scalar $1+2$ diagram. This is the diagram that produces the slow beat before the horizon entry of the infrared tensor. The upper line is the freely propagating sound wave generated by the first interaction in Fig.~\ref{fig:third}(b).}
\label{fig:thirdbeat}
\end{figure}

The physical picture is now similar to second order.
Two incoming waves first generate a higher order sound wave. The higher order sound wave is a result of forced vibration. The freely propagating part of the higher order sound wave then interacts with an initial wave of nearly the same frequency, leading to a continuous source. 
This continuous source with a slow beat is deep inside the horizon and is almost independent of the enhanced regime $[k_-,k_+]$. The resulting contribution to the third order SIGW spectrum is that $\Omega_{33}$ scales as $k^3\ln^2(k_*/k)$ in RD, and as $k^{3-2b}$ for $0<w<1/3$. The higher order corrections have different amplitudes, but the infrared scaling is the same as at second order.

For completeness, we evaluate the other three diagrams in Fig.~\ref{fig:third} in Appendix~\ref{app:otherthird}. We show that the other three diagrams do not give a stronger infrared scaling.

\section{The same slow beat at arbitrary orders}
\label{sec:all}
At third order, we see that the newly generated sound wave propagates freely and interacts with an initial wave. At higher orders, more interactions can take place before the final interaction. Fig.~\ref{fig:general} shows such interactions. As discussed above, panel (b) of Fig.~\ref{fig:general} is the diagram that contributes to the infrared scaling of SIGWs.

\begin{figure}[htb]
\centering
\includegraphics[trim=2pt 2pt 2pt 2pt,clip,width=\linewidth]{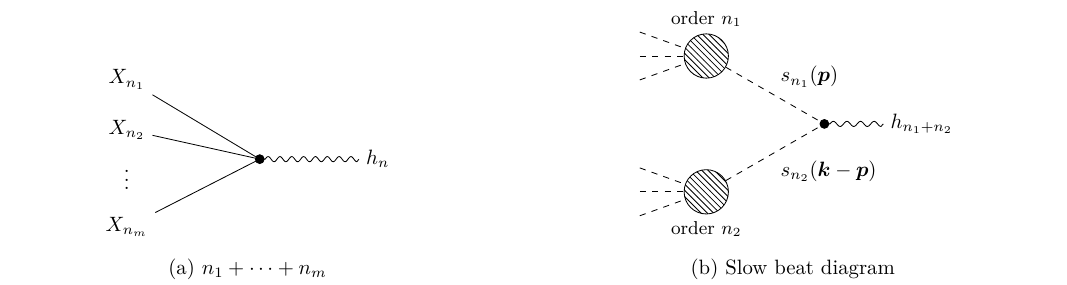}
\caption{Diagrams for the induced higher order tensor modes. In (a), $X_{n_a}$ can be a scalar, vector, or tensor perturbation. In (b), each shaded region contains the earlier interactions that generate a sound wave.}
\label{fig:general}
\end{figure}

Based on our notations, the equation of motion for the $n$-th order Bardeen potential takes the form
\begin{equation}
 \Psi_n''(\bm p)+\frac{4+2b}{\eta}\Psi_n'(\bm p)+c_s^2p^2\Psi_n(\bm p)
  =\mathcal S_n^{\rm scalar}(\bm p,\eta;\zeta),
\label{eq:n-eom}
\end{equation}
Here $\mathcal S_n^{\rm scalar}$ denotes the full source of order $n$. Its expansion contains $n$ initial scalar amplitudes and the corresponding momentum convolutions. The right hand side of Eq.~(\ref{eq:n-eom}) could be very complicated. However, we do not need to derive the source term. Instead, we only need to focus on the homogeneous solution, where $\Psi_n$ has the freely propagating frequency $c_sp=c_s k_* v$. And this homogeneous solution drives the $n$-th order velocity perturbation in the form:
\begin{equation}
  \mathcal V_{ni}^{\text{prop}}(\bm p,\eta)
 =-\frac{2ip_i}{3(1+w)\cH^2}
       [\Psi_n'+\cH\Phi_n](\bm p,\eta).
\label{eq:nvelocityconstraint}
\end{equation}
This process generates an $n$-th order velocity whose envelope is $\eta^{-b}$, just as at first order. For momenta $\bm p$ and $\bm k-\bm p$, two such waves therefore contribute to the $n$-th order source term in the form
\begin{equation}
 a^2(\bar\rho+\bar P)
 \mathcal V_{n_1}^{\rm prop}\mathcal V_{n_2}^{\rm prop}
 \ \supset\ 
 \eta^{-2-2b}
 \cos[c_sk_*(v-u)\eta+\Delta\varphi],
 \label{eq:higherbeat}
\end{equation}
The slow beat frequency is a result of the convolution of $V_{n_1}^{\rm prop}\mathcal V_{n_2}^{\rm prop}$ and is almost frozen before the horizon entry of the infrared tensor. Thus the source is only affected by the expansion of the background and is nearly independent of the enhanced regime $[k_-, k_+]$. Finally, this results in $k^{3-2b}$ scaling for $b>0$, and logarithmic scaling for $b=0$.

For higher orders, $n\ge 4$, there is another diagram that can also generate the slow beat, as shown in Fig.~\ref{fig:tensorbeat}. Two tensors of nearly equal wavenumber have a difference frequency $k_*(v-u)=\mathcal{O}(k)$ and can contribute to the infrared scaling of SIGWs. To see this explicitly, consider the freely propagating parts of the two generated tensors, with $p,|\bm k-\bm p|\gg k$. Their oscillations have the form
\begin{equation}
 h_{2,ab}^{\rm prop}(\bm p,\eta)
 =\left(\frac{\eta}{\eta_s}\right)^{-1-b}
 \left[A_{ab}(\bm p)\cos(p\eta)+B_{ab}(\bm p)\sin(p\eta)\right],
 \label{eq:tensorcarrier}
\end{equation}
where $A_{ab}$
and $B_{ab}$ are time independent coefficients induced by lower order perturbations. The stress carried by such an interaction can be obtained as \cite{Isaacson:1968zza}:
\begin{equation}
 \Pi_{ij}^{hh}
 =\frac{1}{32\pi Ga^2}
 \left\langle\partial_i h_{2,ab}^{\rm prop}
                 \partial_j h_{2,ab}^{\rm prop}\right\rangle\,
 \qquad S_{4,\lambda}^{hh}=16\pi Ga^2e_\lambda^{ij}\Pi_{ij}^{hh}.
 \label{eq:tensorstress}
\end{equation}
The product of the two tensor modes becomes a convolution in Fourier space, with wavelengths $\bm p$ and $\bm k-\bm p$, leading to a source term 
\begin{equation}
 S_{4,\lambda}^{hh,\rm beat}(\bm k,\eta)
 =\frac14\left(\frac{\eta}{\eta_s}\right)^{-2-2b}
 \int\frac{\dd^3p}{(2\pi)^3}e_\lambda^{ij}(\bm k)p_ip_j
 \left[\tilde{A}\cos(\Delta\omega_T\eta)
       +\tilde{B}\sin(\Delta\omega_T\eta)\right].
 \label{eq:tensorbeatsource}
\end{equation}
This is just the slow beat as discussed above. The envelope of the source is $a^{-2}\propto \eta^{-2-2b}$. Consequently, Fig.~\ref{fig:tensorbeat} reproduces the  time integral in Eq.~(\ref{eq:ir-beat-response}), giving the same infrared scaling $k^{3-2b}$ for $b>0$ and logarithmic scaling for $b=0$. 
\begin{figure}[htb]
\centering
\includegraphics[trim=2pt 2pt 2pt 2pt,clip]{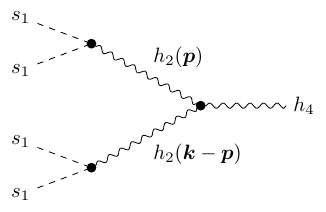}
\caption{A tensor--tensor beat, illustrated at its lowest order by $h_2+h_2\to h_4$.
The same interaction occurs for $h_{n_1}+h_{n_2}\to h_n$ with $n_1,n_2\ge2$ and $n=n_1+n_2\ge4$. Each freely propagating tensor redshifts as $a^{-1}$, so the source envelope is again $a^{-2}\propto\eta^{-2-2b}$ and the difference frequency remains $k_*(v-u)=\mathcal O(k)$.}
\label{fig:tensorbeat}
\end{figure}

However, readers should keep in mind that the dominant slow beat contribution accumulates between the reference time $\eta_s$ and horizon entry of the infrared tensor, namely
\begin{equation}
   \eta_s < \eta< k^{-1}.
\label{eq:window}
\end{equation}
If the primordial power spectrum is broad and flat instead of a bump in the regime $[k_-, k_+]$, the timescale for the slow beat to accumulate can be negligible. As an extreme example, consider a scale invariant primordial power spectrum $\mathcal{P}_\zeta=\mathcal{O}(1)$ during RD. Then $\Omega_{\rm GW}$ is also a constant. There is always a mode $k$ in the power spectrum that satisfies $k^{-1}\simeq \eta_s$ so the slow beat does not have time to accumulate. Therefore, the scalar waves do not have enough time to record the expanding background and the infrared scaling is no longer dominated by the background.

\section{Non-Gaussian initial perturbations}
\label{sec:nongaussian}
The preceding slow beat argument concerns time evolution and does not require Gaussian initial statistics. In the presence of primordial non-Gaussianities, the above discussions are still valid. This is because non-Gaussianities do not change the sound speed or the freely propagating part of the perturbations. In other words, the Green's function of the perturbations remains unchanged. The origin of the slow beat is therefore the same for Gaussian and non-Gaussian initial perturbations.

Non-Gaussian statistics can be generated during inflation and can substantially change the PBH abundance inferred from a given scalar power spectrum~\cite{Sasaki:2018dmp}. Their effect on SIGW amplitudes and spectral features has been studied for local primordial non-Gaussianity~\cite{Nakama:2016gzw,Cai:2018dig,Unal:2018yaa,Yuan:2020iwf,Adshead:2021hnm,Garcia-Saenz:2022tzu,Li:2023qua,Wang:2023ost,Yuan:2023ofl,Li:2023xtl,Inui:2023qsd,Perna:2024ehx,Inui:2024fgk,Li:2025met,Zeng:2025cer,Perna:2026szd,Liu:2023ymk}. This changes the statistical weights of the momentum configurations, rather than the propagation speed entering their beat frequency.

For instance, a local type non-Gaussian model is studied in \cite{Yuan:2023ofl}:
\begin{equation}
 \zeta=\zeta_g+F_{\rm NL}
       \left(\zeta_g^2-\langle\zeta_g^2\rangle\right)
       +G_{\rm NL}\zeta_g^3,
 \qquad F_{\rm NL}=\frac35f_{\rm NL},\quad
 G_{\rm NL}=\frac9{25}g_{\rm NL},
 \label{eq:nglocal}
\end{equation}
where the authors calculated the full second order SIGW spectrum in RD for this non-Gaussian model. The resulting $\Omega_{\rm GW}$ contains coefficients up to $G_{\rm NL}^4$, as well as $F_{\rm NL}^2$, $F_{\rm NL}^4$, and mixed contributions such as $F_{\rm NL}^2G_{\rm NL}^2$. For a finite enhanced band, the authors reproduced the infrared scaling $k^3\ln^2(k_*/k)$ for all the non-Gaussian diagrams.

The physical extension to the constant $w$ background considered here follows from the accumulation interval. For perturbations in the enhanced regime, the oscillation of the slow beat is negligible throughout $\eta_s<\eta< k^{-1}$, within the same constant $w$ era. Non-Gaussian correlations do not change either the envelope $\eta^{-2-2b}$ or the slow beat frequency $c_sk_*(v-u)=\mathcal O(k)$. Since they do not add a rapid time oscillation that would interrupt the slow beat, the waves still have time to record the expansion of the background through Eq.~(\ref{eq:ir-explicit-time-response}).
The extension to primordial non-Gaussianity discussed here is based on a physical interpretation rather than a complete calculation, and a systematic calculation and numerical validation of its infrared scaling for general constant $w$ are left for future work.

\section{Conclusions}
\label{sec:conclusions}
The infrared spectrum of SIGW can be understood through the slow beat of rapidly oscillating waves. At second order, the two velocities in the fluid stress have nearly equal frequencies.
Their convolution generates a slow beat which accumulates from the initial time to tensor horizon entry, $\eta_s<\eta<k^{-1}$.
During this period, the sound waves are deep inside the horizon and the oscillation of the slow beat is negligible. As a result, the slow beat source accumulates over time, recording information about the expansion of the background. The infrared scaling for $\Omega_{\rm GW}$ is thus $k^{3-2b}$ for $0<w<1/3$ and $k^3\ln^2(k_*/k)$ at $w=1/3$.

At third order, two scalar perturbations generate a sound wave whose freely propagating frequency is nearly the same as that of the remaining first order wave. This creates the same slow beat described above. Higher order diagrams work in a similar way. The homogeneous solution of $\Psi$ contributes to the freely propagating part of the velocity perturbation and finally creates a slow beat source.  A broad power spectrum will not destroy such a picture as long as there is enough time for the waves to record the expanding background before tensor horizon entry, $\eta_s<\eta<k^{-1}$.

Non-Gaussianity does not remove this distinction as it neither causes a new rapid oscillation frequency nor changes the speed of sound waves.

In summary, we demonstrate that the infrared scaling of SIGWs can be a probe of the cosmological background even when arbitrary order corrections or non-Gaussianities are considered.

\begin{acknowledgments}
C.Y. acknowledges the assistance of OpenAI GPT-6 Astra in preparing the schematic illustration in Fig.~\ref{fig:slow-beat-mechanism}. The author directed its preparation, reviewed and revised the output, and takes responsibility for the final illustration.
C.Y. acknowledges the use of xPand~\cite{Pitrou:2013hga} for cosmological perturbation expansions.
\end{acknowledgments}

\bibliographystyle{apsrev4-1}
\bibliography{./ref.bib}
\clearpage
\onecolumngrid
\appendix

\section{Details of the Second order SIGWs}
\label{app:second}
\subsection{Explicit expression of Eq.~(\ref{eq:beatsource})}
For completeness, the time independent amplitude introduced in Eq.~(\ref{eq:beatsource}) is
\begin{align}
  B_\lambda(\bm k,\bm p)
 &=\frac{3(1+w)2^{2b+6}\Gamma^2(b+5/2)}
 {\pi(5+3w)^2(1+b)^2}(c_sk_*)^{-2-2b}
 \frac{e_\lambda^{ij}p_ip_j}{(uv)^{1+b}}
 \zeta(\bm p)\zeta(\bm k-\bm p).
 \label{eq:beatnormalization}
\end{align}
It follows directly by substituting Eq.~(\ref{eq:velocityenvelope}) into Eq.~(\ref{eq:velocitysource}) and taking the first term of Eq.~(\ref{eq:secondtrig}).

\subsection{Transfer functions and spectrum of SIGWs}
The following expressions connect the velocity source with the conventional transfer function calculation~\cite{Kohri:2018awv,Gong:2019mui}.
For the transfer functions in Eq.~(\ref{eq:regulartransfer}), define $T_u\equiv T_w(uk_*,\tau)$ and $T_v\equiv T_w(vk_*,\tau)$. Then the transfer function of the source can be written as
\begin{align}
  f_w(u,v;\tau)&=T_uT_v+\frac{2}{3(1+w)}
 \left(T_u+\frac{T_u'}{\cH}\right)
 \left(T_v+\frac{T_v'}{\cH}\right),\label{eq:fullsource}\\
\mathcal I_w(u,v,\tilde{k};\eta)&=2k^2\int_0^\eta\dd\tau\,
 G_k(\eta,\tau)f_w(u,v;\tau),\\
 h_{2,\lambda}(\bm k,\eta)&=2\int\frac{\dd^3p}{(2\pi)^3}
 \frac{e_\lambda^{ij}p_ip_j}{k^2}\,
 \zeta(\bm p)\zeta(\bm k-\bm p)\mathcal I_w .
 \label{eq:fullresponse}
\end{align}
$G_k$ is the retarded Green function of Eq.~(\ref{eq:tensorgreen}). Our Fourier convention is
\begin{equation*}
 h_{2,ij}(\bm x,\eta)=\sum_{\lambda=+,\times}
 \int\frac{\dd^3k}{(2\pi)^3}\,
 e_{ij}^{\lambda}(\bm k)h_{2,\lambda}(\bm k,\eta)
 e^{i\bm k\cdot\bm x}.
\end{equation*}
For statistically homogeneous and isotropic perturbations with equal, uncorrelated polarization powers, the equal time two point function is
\begin{equation*}
 \left\langle h_{2,\lambda}(\bm k,\eta)
 h_{2,\lambda'}^*(\bm k',\eta)\right\rangle
 =(2\pi)^3\delta^3(\bm k-\bm k')\delta_{\lambda\lambda'}
 \frac{2\pi^2}{k^3}\frac{\mathcal P_{h_2}(k,\eta)}{2}.
\end{equation*}
With our unit normalized polarization tensors, contraction of the tensor indices gives
\begin{align*}
 \left\langle h_{2,ij}(\bm k,\eta)h_2^{ij}(\bm k',\eta)\right\rangle
 &=(2\pi)^3\delta^3(\bm k+\bm k')
 \frac{2\pi^2}{k^3}\mathcal P_{h_2}(k,\eta),\\
 \mathcal P_{h_2}(k,\eta)
 &=\frac{k^3}{2\pi^2}\left\langle
 h_{2,ij}(\bm k,\eta)h_2^{ij}(-\bm k,\eta)\right\rangle'.
\end{align*}
Here the prime means that the factor $(2\pi)^3\delta^3(\bm k+\bm k')$ is removed before setting $\bm k'=-\bm k$. Thus $\mathcal P_{h_2}$ is the dimensionless power summed over both polarizations and satisfies $\langle h_{2,ij}(\bm x,\eta)h_2^{ij}(\bm x,\eta)\rangle =\int\dd\ln k\,\mathcal P_{h_2}(k,\eta)$. Our convention for the scalar power spectrum is
\begin{equation}
    \langle\zeta(\bm p)\zeta(\bm p')\rangle
=(2\pi)^3\delta^3(\bm p+\bm p')\,2\pi^2\Pz(p)/p^3.
\end{equation}
For Eq.~(\ref{eq:fullpower}), we introduce the momentum ratios $\tilde v\equiv p/k=v/\tilde k$ and $\tilde u\equiv|\bm k-\bm p|/k=u/\tilde k$. Follow the semi-analytical method in \cite{Kohri:2018awv}, the power spectrum of the second order tensor mode is 
\begin{align}
 \overline{\mathcal P_h(k,\eta)}
 &=\frac{4}{(k\eta)^2}\int_0^\infty\dd\tilde v
 \int_{|1-\tilde v|}^{1+\tilde v}\dd\tilde u\,
 \left[\frac{4\tilde v^2-(1+\tilde v^2-\tilde u^2)^2}{4\tilde v\tilde u}\right]^2
 \overline{\widehat I_w^{\,2}(\tilde v,\tilde u,k\eta)}\Pz(k\tilde v)\Pz(k\tilde u).
 \label{eq:fullpower}
\end{align}
Here $\widehat I_w(\tilde v,\tilde u,k\eta)\equiv(k\eta)I_w(\tilde v,\tilde u,k\eta)$ and $I_w(\tilde v,\tilde u,k\eta)=\mathcal I_w(\tilde k\tilde u,\tilde k\tilde v,\tilde k;\eta)$.
The overline averages rapid tensor oscillations. We extract the factor $k \eta$ from $I_w$ so that the rest kernel function $I_w$ would approach constant in the sub-horizon limit $k\eta\gg 1$ \cite{Kohri:2018awv}. And this $\eta$ dependence is canceled with $(k/\mathcal{H})^2$ in Eq.~(\ref{eq:ir-omega-definition}) so that the $\eta$ dependence of Eq.~(\ref{eq:ir-omega-definition}) only comes from the factor $k^3$ and $h_{ij}$.

\subsection{Checking the remaining diagrams of the second order SIGWs}
\label{sec:secondaudit}
Eq.~(\ref{eq:fullsource}) can be rewritten as
\begin{equation}
 f_w=(1+c_w)T_vT_u
 +c_w\left(T_v\frac{T_u'}{\cH}+\frac{T_v'}{\cH}T_u\right)
 +c_w\frac{T_v'T_u'}{\cH^2},
 \qquad c_w=\frac{2}{3(1+w)}.
 \label{eq:secondcomplete}
\end{equation}
Now we have three terms in total and they correspond to the three diagrams in Table~\ref{tab:secondcensus} respectively.  Let $\theta_v^\Phi=c_svk_*\eta-\pi(b+2)/2$ and $\theta_u^\Phi=c_suk_*\eta-\pi(b+2)/2$. Eq.~(\ref{eq:potentialenvelope}) then gives
\begin{align}
 T_v&\simeq\mathcal N_w(c_svk_*\eta)^{-2-b}\cos\theta_v^\Phi,
 \\
 T_v'&\simeq-\mathcal N_w c_svk_*(c_svk_*\eta)^{-2-b}
 \sin\theta_v^\Phi,\\
 \frac{T_v'}{\cH}&\simeq
 -\frac{\mathcal N_w}{1+b}(c_svk_*\eta)^{-1-b}\sin\theta_v^\Phi.
 \label{eq:secondderivative}
\end{align}
As a result, the envelopes for the three terms in Eq.~(\ref{eq:fullsource}) can be obtained. Table~\ref{tab:secondcensus} summarizes the time dependence of these three sources.

\begin{table}[htb]
\caption{The full diagrams and the envelope of the second order source.}
\label{tab:secondcensus}
\centering\small
\renewcommand{\arraystretch}{1.15}
\begin{tabular}{@{}ll@{}}
\toprule
\sourcecell{.30\linewidth}{Diagram}
&\sourcecell{.63\linewidth}{Leading subhorizon source time dependence}\\
\midrule
\sourcecell{.30\linewidth}{\centering\raisebox{-.5\height}{\includegraphics[trim=2pt 2pt 2pt 2pt,clip]{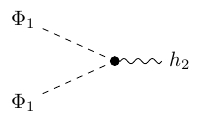}}}
&\sourcecell{.63\linewidth}{$\eta^{-4-2b}\cos\theta_v^\Phi\cos\theta_u^\Phi$.}\\[6pt]
\sourcecell{.30\linewidth}{\centering\raisebox{-.5\height}{\includegraphics[trim=2pt 2pt 2pt 2pt,clip]{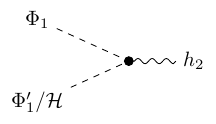}}}
&\sourcecell{.63\linewidth}{$\eta^{-3-2b}\cos\theta_v^\Phi\sin\theta_u^\Phi+(v\leftrightarrow u).$}\\[6pt]
\sourcecell{.30\linewidth}{\centering\raisebox{-.5\height}{\includegraphics[trim=2pt 2pt 2pt 2pt,clip]{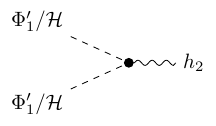}}}
&\sourcecell{.63\linewidth}{$\eta^{-2-2b}\sin\theta_v^\Phi\sin\theta_u^\Phi$.}\\
\bottomrule
\end{tabular}
\end{table}

The sources are weighted by $\eta d\eta$ and accumulate in time to generate the second order tensor mode through time integrals of the form $\int S~\eta d\eta$. Consequently the first two diagrams in the table decay faster in time compared with the third diagram. In RD, the third diagram produces the logarithmic scaling, dominating the infrared scaling of SIGWs.

\section{The induced scalar and velocity at second order}
\label{app:scalar}
\subsection{The explicit second order scalar source}
Consider two linear scalar modes with momenta $\bm q$ and $\bm p-\bm q$ that generate a second order scalar mode. Let
\begin{equation}
 q=rk_*,\qquad |\bm p-\bm q|=\ell k_*,\qquad p=vk_*,
 \qquad
 T_r=T_w(rk_*,\eta),\quad T_\ell=T_w(\ell k_*,\eta).
 \label{eq:v9ratios}
\end{equation}
The transfer function includes the conversion from the initial curvature perturbation,  $\Phi_1(\bm q,\eta)=T_r\zeta(\bm q)$.

The difference of the two Bardeen potentials at second order is fixed by the traceless part of the  Einstein equation:
\begin{equation}
 (\Phi_2-\Psi_2)(\bm p,\eta)
 =\int\frac{\dd^3q}{(2\pi)^3}
 \left[2T_rT_\ell-\mathcal M\mathcal B_w\right]
 \zeta(\bm q)\zeta(\bm p-\bm q),
 \label{eq:v9slip}
\end{equation}
where the angular part and the transfer function are:
\begin{align}
 \mathcal M(v,r,\ell)
 &=\frac{v^4+2v^2(r^2+\ell^2)-3(r^2-\ell^2)^2}{8v^4},
 \label{eq:v9M}\\
 \mathcal B_w(r,\ell;\eta)
 &=\frac{2(5+3w)}{3(1+w)}T_rT_\ell
 +\frac{4}{3(1+w)\cH}(T_r'T_\ell+T_rT_\ell')
 +\frac{4}{3(1+w)\cH^2}T_r'T_\ell'.
 \label{eq:v9B}
\end{align}
The scalar evolution equation is
\begin{equation}
 \Psi_2''+\frac{4+2b}{\eta}\Psi_2'+wp^2\Psi_2
 =\int\frac{\dd^3q}{(2\pi)^3}
 Q_w(v,r,\ell;\eta)\zeta(\bm q)\zeta(\bm p-\bm q),
 \label{eq:v9forced}
\end{equation}
where the source is given by
\begin{align}
 Q_w={}&-k_*^2\left[
 \frac{3w-1/3}{4}(v^2-r^2-\ell^2)
 +2w(r^2+\ell^2)\right]T_rT_\ell
 +\frac{1+3w}{2}T_r'T_\ell'\nonumber\\
 &+\mathcal M\left(\cH\mathcal B_w'
                  -\frac{v^2k_*^2}{3}\mathcal B_w\right)-\frac{(1/3-w)k_*^2(v^2-r^2-\ell^2)}
 {3(1+w)\cH^2}
 (T_r'+\cH T_r)(T_\ell'+\cH T_\ell).
 \label{eq:v9Q}
\end{align}
This expression includes the constraint for $\Phi_2-\Psi_2$.

\subsection{The second order velocity}
Here, we do not consider the initial value of $\Psi_2$ so that the solution of $\Psi_2$ is given by
\begin{align}
 \Psi_2(\bm p,\eta)
 &=\int\frac{\dd^3q}{(2\pi)^3}
 K_\Psi(v,r,\ell;\eta)\zeta(\bm q)\zeta(\bm p-\bm q),\\
 K_\Psi(v,r,\ell;\eta)
 &=\int_{\eta_i}^{\eta}\dd\tau\,
 G_\Psi(p;\eta,\tau)Q_w(v,r,\ell;\tau).
 \label{eq:v9retarded}
\end{align}
The scalar Green function is defined by
\begin{equation}
 \left[\partial_\eta^2+\frac{4+2b}{\eta}\partial_\eta
       +c_s^2p^2\right]G_\Psi(p;\eta,\tau)
 =\delta(\eta-\tau),\qquad G_\Psi=0\quad(\eta<\tau).
\end{equation}
The explicit expression is
\begin{equation}
 G_\Psi(p;\eta,\tau)=\frac\pi2\Theta(\eta-\tau)
 \frac{\tau^{\alpha+1}}{\eta^\alpha}
 \left[J_\alpha(c_sp\tau)Y_\alpha(c_sp\eta)
       -Y_\alpha(c_sp\tau)J_\alpha(c_sp\eta)\right].
 \label{eq:v9Gs}
\end{equation}

On the other hand, the 0i component of the Einstein equation gives the second order velocity. Let $\mathsf L_i{}^j=\partial_i\partial^j/\nabla^2$ denote the longitudinal projection and let $\delta_1=\delta\rho_1/\bar\rho$. The second order $0i$ equation is then
\begin{align}
 \partial_i(\Psi_2'+\cH\Phi_2)
 +\mathsf L_i{}^j\left[(\Phi_1'-4\cH\Phi_1)
                       \partial_j\Phi_1\right]=-4\pi Ga^2(\bar\rho+\bar P)
 \left\{\mathcal V_{2i}^{\rm S}
       +\mathsf L_i{}^j[(\delta_1-3\Phi_1)\mathcal V_{1j}]\right\}.
 \label{eq:v9momentumfull}
\end{align}
Solving for the velocity gives
\begin{align}
 \mathcal V_{2i}^{\rm S}={}
 -\frac{2}{3(1+w)\cH^2}\partial_i(\Psi_2'+\cH\Phi_2)-\frac{2}{3(1+w)\cH^2}
 \mathsf L_i{}^j[(\Phi_1'-4\cH\Phi_1)\partial_j\Phi_1]
 -\mathsf L_i{}^j[(\delta_1-3\Phi_1)\mathcal V_{1j}].
 \label{eq:v9velocityfull}
\end{align}
The first order density contrast is fixed by the $00$ equation,
\begin{equation}
 \delta_1=-2\Phi_1-\frac{2}{\cH}\Phi_1'
          +\frac{2}{3\cH^2}\nabla^2\Phi_1.
\end{equation}
The last two terms in Eq.~(\ref{eq:v9velocityfull}) define $\mathcal R_{2i}$ in Eq.~(\ref{eq:velocityconstraint}). They contain the contribution of the first order scalar modes.

\section{Comparison of the remaining third order sources}
\label{app:otherthird}
In the main text, we study the propagating acoustic part of the scalar $1+2$ diagram. In this section, we compare the time accumulation of the other contributions with the scalar $1+2$ diagram. For all the third order diagrams, we use
\begin{equation}
 q=r k_*,\quad |\bm p-\bm q|=\ell k_*,\quad
 p=v k_*,\quad |\bm k-\bm p|=u k_*.
 \label{eq:otherrouting}
\end{equation}
The three initial magnitudes $r k_*$, $\ell k_*$, and $u k_*$ fall in the enhanced band $[k_-,k_+]$. The intermediate magnitude satisfies $|v-u|k_*\leq k$. 

\subsection{The third order tensor mode}
At fixed convolution momenta well inside the sound horizon,
\begin{equation}
 \Phi_1\propto\eta^{-2-b},\qquad
 \delta_1\propto\eta^{-b},\qquad
 \mathcal V_1\propto\eta^{-b},\qquad
 a^2\bar\rho\propto\eta^{-2}.
 \label{eq:otherlinearcount}
\end{equation}
These expressions are useful in analyzing the envelopes. For a tensor still outside  the horizon, the solution of the third order tensor mode takes the form
\begin{equation}
 h_3(\eta)
 =\frac{1}{1+2b}\int_{\eta_s}^{\eta}\dd\tau\,\tau S_3(\tau)
 \left[1-\left(\frac{\tau}{\eta}\right)^{1+2b}\right],
 \qquad k\eta\ll1.
 \label{eq:othergreenweight}
\end{equation}
Below, we will discuss different sources $S_3$.

\subsection{The \texorpdfstring{$1+1+1$}{1+1+1} diagram}
We are interested in the slowest term in the third order source. Using Eq.~(\ref{eq:otherlinearcount}) to analyze the third order sources, the slowest term we find is 
\begin{equation}
 S_3^{(111)}\supset
 a^2\bar\rho\,\delta_1\mathcal V_{1i}\mathcal V_{1j}
 \propto\eta^{-2-3b}.
 \label{eq:otherdirectenvelope}
\end{equation}
All the possible frequencies are 
\begin{equation}
 \omega_{3}=c_sk_*(\pm r\pm \ell\pm u).
\label{eq:othertriplefreq}
\end{equation}

Let $S_3^{(111)}=S_{111}(\eta/\eta_s)^{-2-3b}$, where $S_{111}$ is a time independent amplitude.
Eq.~(\ref{eq:othergreenweight}) then implies:
\begin{align}
 |h_3^{(111)}(\eta)|
 \leq\frac{S_{111}\eta_s^2}{1+2b}
 \int_1^{\eta/\eta_s}\dd t \,t^{-1-3b}
 =\frac{S_{111}\eta_s^2}{1+2b}\times \begin{cases}
 (1-(\eta/\eta_s)^{-3b})/(3b),&b>0,\\
 \ln(\eta/\eta_s),&b=0.
 \end{cases}
 \label{eq:accumulationbound}
\end{align}
The propagating scalar $1+2$ diagram has an envelope $\sim \eta^{-2-2b}$. Consequently, throughout $0\leq b<1$, the scaling of $h_3^{(111)}$ is at most $\sim\ln k$ during RD and it can never exceed logarithmic growth.  Thus the $1+1+1$ diagram cannot produce a scaling stronger than the scalar $1+2$ diagram.

\subsection{The forced part of the scalar \texorpdfstring{$1+2$}{1+2} diagram}
Besides free propagation, there are two frequencies $c_sk_*(r+\ell)$ and $c_sk_*(r-\ell)$ from the forced vibration. 
Write $\omega_p=c_sk_*v$ and $\omega_d=c_sk_*(r\pm\ell)$. Then the slowest damping of the second order velocity is
\begin{equation}
  \mathcal V_2^{\rm forced}
 \propto \left({\eta\over\eta_s}\right)^{-2b}\frac{e^{i\omega_d\eta}}
 {\omega_p^2-\omega_d^2}.
 \label{eq:forcedvelocityenvelope}
\end{equation}
The two driving waves provide the envelope $\eta^{-2b}$. The source term then behaves as
\begin{equation}
 S_3^{(12),\rm forced}
  \propto\eta^{-2}\eta^{-b}\eta^{-2b}e^{i\omega_3\eta}
 =\eta^{-2-3b}e^{i\omega_3\eta},
\label{eq:otherforcedscalar}
\end{equation}
or faster. Thus the scaling in the infrared regime cannot be stronger than logarithmic growth. 

\subsection{The vector \texorpdfstring{$1+2$}{1+2} diagram}
\label{app:vectorcheck}
Let $V_{2i}$ denote the second order vector perturbation, with
$g_{0i}=a^2V_{2i}$ and $\partial_iV_{2i}=0$.
This is distinct from the fluid velocity perturbation $\mathcal V_{2i}$. The vector perturbation $V_{2i}$ obeys
\begin{equation}
 V_{2i}'+2\cH V_{2i}=Q_{Vi},\qquad
 V_{2i}(\eta)=a^{-2}(\eta)
 \left[a^2(\eta_s)V_{2i}(\eta_s)
       +\int_{\eta_s}^{\eta}\dd\tau\,a^2(\tau)Q_{Vi}(\tau)\right].
 \label{eq:vectorresponse}
\end{equation}
Taking the transverse divergence of the spatial Einstein equation gives the explicit source:
\begin{equation}
  Q_{Vi}=-\nabla^{-2}\mathcal T_i{}^j\partial^l
 \left[4\partial_j\Phi_1\partial_l\Phi_1
 +6(1+w)\cH^2\mathcal V_{1j}\mathcal V_{1l}\right],
 \qquad \mathcal T_i{}^j=\delta_i{}^j-\frac{\partial_i\partial^j}{\nabla^2}.
 \label{eq:vectorfullsource}
\end{equation}
In Fourier space, set
$q_i^\perp=(\delta_i{}^j-p_i p^j/p^2)q_j$. Then
\begin{align}
 Q_{Vi}(\bm p,\eta)&=4i\int\frac{\dd^3q}{(2\pi)^3}
 \mathcal P_i(\bm p,\bm q)
 \left[T_rT_\ell+\frac{2(T_r'+\cH T_r)(T_\ell'+\cH T_\ell)}{3(1+w)\cH^2}\right]
 \zeta(\bm q)\zeta(\bm p-\bm q),
 \label{eq:vectorFouriersource}\\
 \mathcal P_i(\bm p,\bm q)&=
\frac{\mathcal T_i{}^j}{2p^2}
 \left[q_j(\bm p\cdot\bm q)
 +(p-q)_j\bm p\cdot(\bm p-\bm q)\right]
 =\frac{q_i^\perp k_*^2}{2p^2}(r^2-\ell^2).
 \label{eq:vectorprojection}
\end{align}
Here $\mathcal P_i$ denotes the time independent coefficients from the projection. 
Except for the driven frequencies, $c_sk_*(r\pm\ell)$, $V_{2i}$ has no freely propagating frequency.
The leading term, $(T_r'+\cH T_r)(T_\ell'+\cH T_\ell)/\cH^2$, has an envelope $\eta^{-2-2b}$ and the corresponding time integral is given by
\begin{equation}
 \begin{aligned}
 \int_{\eta_s}^{\eta}\dd\tau\,a^2Q_{Vi}\sim  k_*^2(r^2-\ell^2)\int_{\eta_s}^{\eta}\dd\tau\,
                  \cos[c_sk_*(r-\ell)\tau]=\frac{k_*(r+\ell)}{c_s}
 \left\{\sin[c_sk_*(r-\ell)\eta]
       -\sin[c_sk_*(r-\ell)\eta_s]\right\}\le{\frac{2k_*(r+\ell)}{c_s}} .
 \end{aligned}
 \label{eq:vectorfinite}
\end{equation}
Unlike the scalar $1+2$ diagram, this time integral has a coefficient $r^2-\ell^2$ so that the result is bounded. This yields the result that $V_{2i}$ scales as $\eta^{-2-2b}$. Then we obtain the envelope of the slowest source terms, such as
\begin{equation}
 S_3^{(12),V}\supset
 \cH^{-2}\Phi_1'\partial_{(i}\nabla^2V_{2j)}
 \propto\eta^2\eta^{-2-b}\eta^{-2-2b}
 =\eta^{-2-3b}.
 \label{eq:vectorsourceenvelope}
\end{equation}
therefore the vector $1+2$ diagram does not provide stronger time accumulation than the propagating part of the scalar $1+2$ source.

\subsection{The tensor \texorpdfstring{$1+2$}{1+2} diagram}
The intermediate tensor has a freely propagating frequency $p$. However, the speed of the tensor mode is the speed of light, instead of $c_s$. Consequently, the induced tensor mode cannot form a beat with the scalar waves. For instance, the scalar--tensor source contains a term such as $\Phi_1\nabla^2h_2$. This term contributes as
\begin{equation}
 S_3^{(12),h}
 \propto\eta^{-3-2b}e^{ik_*(\pm v\pm c_su)\eta},
 \label{eq:othertensor}
\end{equation}
Since the frequency satisfies $|p-c_su k_*|\geq(1-c_s)k_-$ for the infrared tensor mode $k\ll k_-$, the time weighted integral
in Eq.~(\ref{eq:othergreenweight}) is finite. The time integral becomes independent of $k$ and thus cannot produce a stronger infrared scaling than the scalar $1+2$ diagram.
\end{document}